# Unsupervised MRI Super Resolution Using Deep External Learning and Guided Residual Dense Network with Multimodal Image Priors

Yutaro Iwamoto, *Member, IEEE*, Kyohei Takeda, Yinhao Li, Akihiko Shiino, and Yen-Wei Chen, *Member, IEEE*

***Abstract*— Deep learning techniques have led to state-of-the-art image super resolution with natural images. Normally, pairs of high-resolution and low-resolution images are used to train the deep learning models. These techniques have also been applied to medical image super-resolution. The characteristics of medical images differ significantly from natural images in several ways. First, it is difficult to obtain high-resolution images for training in real clinical applications due to the limitations of imaging systems and clinical requirements. Second, other modal high-resolution images are available (e.g., high-resolution T1-weighted images are available for enhancing low-resolution T2-weighted images). In this paper, we propose an unsupervised image super-resolution technique based on simple prior knowledge of the human anatomy. This technique does not require target T2WI high-resolution images for training. Furthermore, we present a guided residual dense network, which incorporates a residual dense network with a guided deep convolutional neural network for enhancing the resolution of low-resolution images by referring to different modal high-resolution images of the same subject. Experiments on a publicly available brain MRI database showed that our proposed method achieves better performance than the state-of-the-art methods.**

Manuscript received xxxxxx. This work was partly supported by JSPS KAKENHI Grant-in-Aid for Early-Career Scientists (18K18078), Grant-in-Aid for Scientific Research (B) (18H03267) and Grant-in-Aid for Challenging Research (Exploratory) (20K21821). The authors would like to thank Enago (www.enago.jp) for the English language review.

Yutaro Iwamoto is with the Department of Engineering Informatics, Faculty of Information and Communication Engineering, Osaka Electro-Communication University, Osaka, Japan (e-mail: yiwamoto@osakac.ac.jp).

Kyohei Takeda and Yinhao Li are with Graduate School of Information Science and Engineering, Ritsumeikan University, Shiga, Japan.

Akihiko Shiino is with Molecular Neuroscience Research Center, Shiga University of Medical Science, Shiga, Japan.

Yen-Wei Chen is with the College of Information Science and Engineering, Ritsumeikan University, Shiga, Japan (e-mail: chen@is.ritsumei.ac.jp, Corresponding Author).

Yutaro Iwamoto and Kyohei Takeda contributed equally to this work.

## I. Introduction

MAGNETIC resonance imaging (MRI) with superior tissue contrast can be employed to visualize detailed internal information of the human body. Multiple MRI sequences (or multimodal MR images) with different tissue contrast such as T1-weighted images (T1WI) and T2-weighted images (T2WI), which are routinely acquired, can provide complementary information. These are widely used together for accurate diagnosis [1]-[3]. The lack of availability of HR images (T2WI) motivates us to use SR techniques to enhance the resolution of T2WI. In clinical applications, such as diagnosis of brain or liver lesions, T1WI provides a high-contrast image of anatomic structure, while T2WI provides a high-contrast image of lesions because the accumulation of water molecules in most lesions results in obviously higher signals than normal tissue [40]. T2WI is generally used more commonly to diagnose lesions. On the other hand, the acquisition times of MRI sequences are different—for e.g., a long time is required for T2WI and a short time for T1WI. Since it is difficult to acquire a high-resolution (HR) image for time-consuming sequences. We usually acquire a high-resolution (fully-sampled) T1WI and a low-resolution (under-sampled) T2WI. Therefore, resolution-enhancement of T2WI is an important issue. Classical interpolation techniques suffer from the problem of image blurring and jaggy artifacts in the interpolated image. To solve this problem, image super resolution (SR) [4]-[22], [33]-[39] has gained attention as a resolution-enhancement technique. SR enhances the resolution based on the relationship between the HR and LR images and prior knowledge of various image characteristics, such as sparse representation [36] and total variation [37].

Recently, deep convolutional neural network (CNN)-based SR methods have been frequently reported in the field of SR for natural images [4]-[8] and medical imaging [13]-[16]. The number of layers used in these methods is increased, and residual [23] and dense [24] structures are employed to prevent overfitting. In medical imaging, three-dimensional (3D) CNN-based SR are proposed to consider the 3D structure of the volume [14]-[16]. In addition, the generative adversarial network generates more realistic images by correcting SR images [6] [15]. As a result, these networks have more parameters and a more complex network structure. Some guided SR techniques are also proposed for enhancing multimodal images, such as the HR panchromatic images and LR multispectral images in remote sensing [20], HR color images and LR depth images in the consumer depth cameras [21] [22], and MRI sequences in medical imaging [17-19]. Unfortunately, these promising methods [4]-[7] [14]-[16] adopt a supervised learning approach and necessitate a large number of ideal HR images. However, in real-world clinical applications, it is generally difficult to acquire a large set of HR medical images because of hardware limitations and clinical requirements.

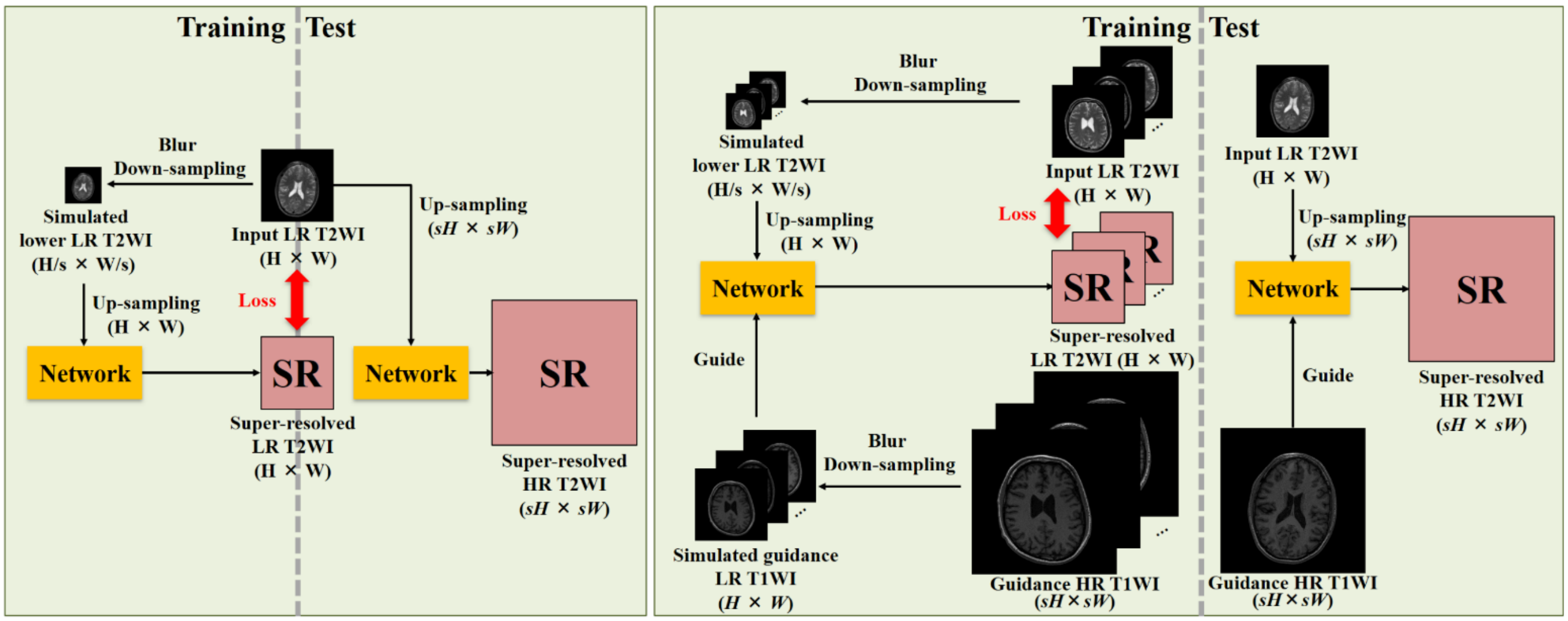

Fig. 1. Unsupervised multimodal guided SR framework. (left) Conventional internal learning-based unsupervised method. (right) Proposed unsupervised method (external learning + guidance by multimodal priors).

A few unsupervised approaches [8]-[13], [33] have been proposed for SR without using HR images for training. These methods are based on self-similarity with the input LR image. Note that the term "unsupervised" refers to the fact that training does not use ideal HR data in this paper. These unsupervised approaches can be divided into two groups. The first group [8]-[10] assumes self-similarity — i.e., similar patches are found across the same and different scales within an image itself, and this group is termed “zero-shot” super resolution (ZSSR) [8]. The test LR image is used as an HR image for training, and its simulated degradation image is used as an LR image for training. Such self-training data is called as internal training data in contrast to the external training data (training data are only used for training). This kind of learning is called internal learning [8]. While external training data-based learning is referred to as external learning. The limitation of internal learning methods is the need to train a model for each set of test data. Therefore most SR architectures only employ simple models (8 hidden layers [8]) and face the problems related to computation time and resources. Further, existing methods [8]-[10] are proposed for natural images. In the second group of methods [11]-[13], in-plane (x-y plane) images of 3D medical imaging are used as HR images, and their simulated degradation images are used as LR images for training. These methods are used to enhance the resolution along the z-axis. In [34, 35], authors proposed cycle GAN-based methods using unpaired LR-HR images for training. Since unpaired LR-HR images are used for training, these methods are also considered as unsupervised SR approaches. In contrast, the conventional supervised SR approaches use paired LR-HR images for trainings. However, these cycle GAN-based methods are useful for real applications, but they need HR images for training. On the other hand, HR images are not available in many real-world applications (e.g., T2WI) due to the limitations of imaging systems and clinical requirements. The lack of availability of HR images motivates us to use only LR images for medical image SR task.

In this paper, we propose an unsupervised SR approach using only LR images for training. We apply deep external learning to improve the accuracy of SR in comparison to the state-of-the-art unsupervised method (ZSSR [8]). Furthermore, we propose a guided deep convolutional neural network for enhancing the resolution of LR images (T2WI) by referring to different modal HR images (T1WI) of the same subject. Note that in our proposed approach, we use only other modal HR images (T1WI) as a guidance. The target modal HR images (T2WI), which are not available, are not used during training. The main contributions are summarized below:

(1) The proposed method uses prepared external LR images as HR images and their simulated degradation images as LR images in the form of external learning to avoid the enormous training time of the test phase in the internal learning. It assumes simple prior knowledge that the anatomical structures of humans are identical. External learning can perform better than internal learning in unsupervised SR of medical images.

(2) We proposed an HR T1WI guided residual dense network to enhance the resolution of LR T2WI. We improve the SR reconstruction accuracy of T2WI in both supervised and unsupervised SR by concatenating the multimodal priors (features of HR T1WI) with the features of LR T2WI.

The rest of this article is organized as follows. Section II describes the proposed method. The results of experiments are presented in Section III. Finally, Section IV outlines conclusion and future work.

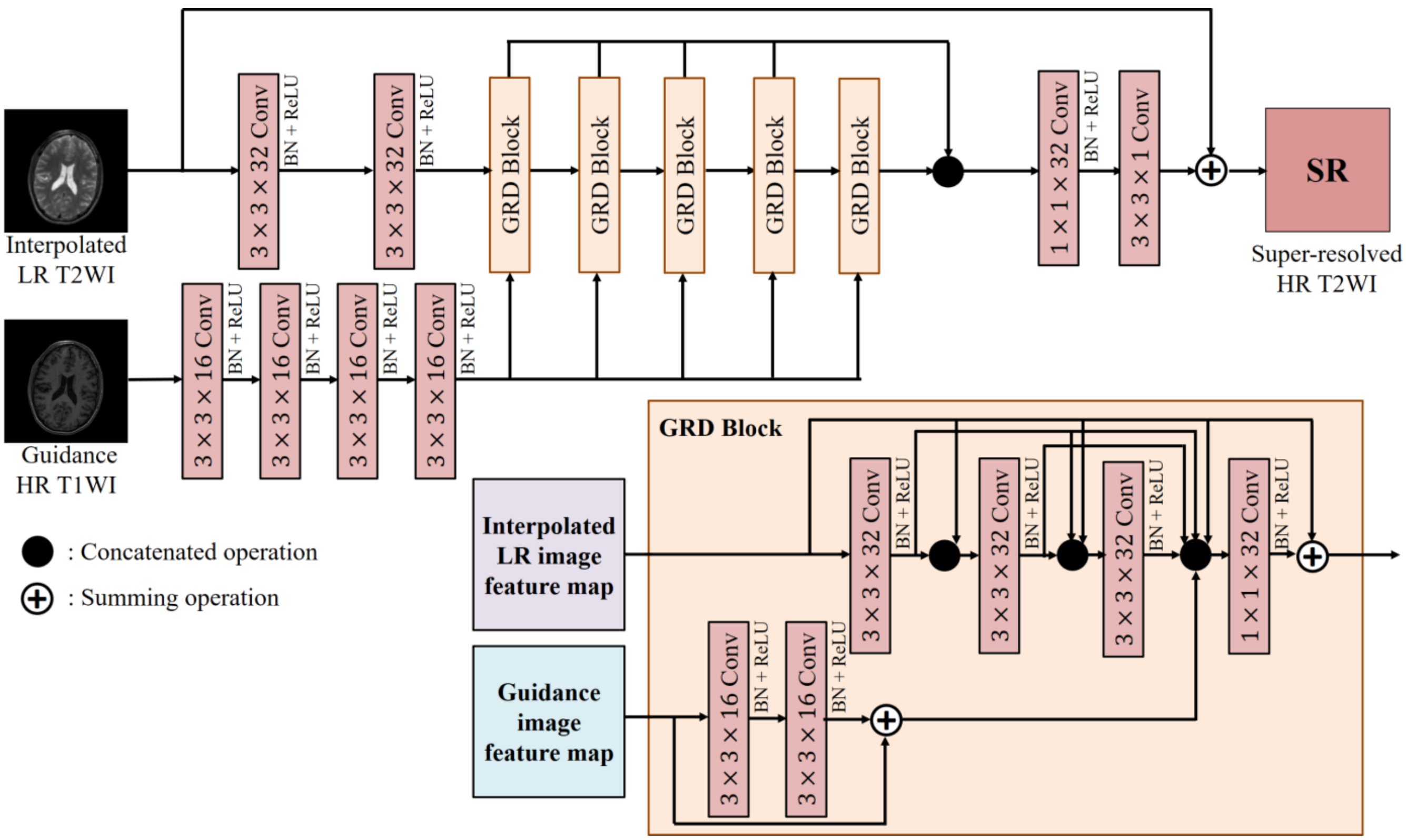

Fig. 2. Network architecture of the proposed method for the resolution enhancement of LR T2WI guided by multimodal priors (HR T1WI). The 3x3x32 Conv means that the kernel size is 3x3 and the number of filters (channels) is 32.

## II. Proposed Method

### *A. Overview*

Figure 1 shows an overview of the proposed method. The method is based on two main concepts. One is the use of external LR training images to train the model. The conventional unsupervised approach trains a model using only a test LR image (internal learning) as shown in Fig. 1 (left). Therefore, the method cannot separate the training and test phases, and a high computational time (including both training and testing) is required. In addition, since deep learning requires numerous training samples for parameter optimization, excessive data augmentation is used to increase the number of training samples from a single LR image. Therefore, we propose to use external LR samples for training as shown in Fig. 1 (right). These images can be easily collected even in real clinical applications. By using external training LR samples, we can separate the training and test phases. Therefore, a lot of time can be spent for training before applying the-state-of-the-art SR framework, and a sufficient number of training samples can be generated without excessive data augmentation. The second concept is to incorporate multimodal priors (T1WI) into the unsupervised SR framework to guide the resolution enhancement of T2WI as shown in Fig. 1 (right). In unsupervised learning, since the network is trained in the LR domain, there is a lack of high-frequency components. By using HR guidance images, we can efficiently enhance the high-frequency information of the test LR image.

### *B. Network architecture*

Figure 2 shows the proposed network architecture. The proposed model is inspired by the residual dense network (RDN) [7], which is the state-of-the-art network for various computer vision tasks such as SR. The model comprises residual and dense structures. The residual network [23] uses shortcut connections, which skip one or more layers. The feature maps are connected by a summing operation. The dense network [24] uses dense connections from any layer to all subsequent layers. The connection is a concatenated operation. The RDN architecture is constructed by some densely connected blocks, and shortcut (skip) connections are arranged outside each dense connection block. The residual training [5] is realized by a shortcut connection from the input to the end of the final layer. The proposed framework can be combined with a wide variety of existing network architectures as a backbone to further enhance their performance. The input of the network is an interpolated LR MRI sequence image (e.g., T2WI). Furthermore, in the proposed method, we add a new branch to incorporate the guided HR T1WI as shown in Fig. 2. The features extracted from the guided HR T1WI (multimodal priors) are concatenated with the features of the target T2WI in newly proposed guided residual dense (GRD) blocks as shown in Fig. 2. Each convolutional layer involves 3 × 3 convolution, batch normalization (BN) [25], and rectified linear unit (ReLU) [26].

### C. Loss function

In the conventional supervised SR, the basic loss [4] is formulated as follows:

$$L_s(\boldsymbol{\Theta}) = \frac{1}{N}\sum_{i=1}^{N}\left\| f\left(D^T \boldsymbol{Y}_{Tr}^i; \boldsymbol{\Theta}\right) - \boldsymbol{X}_{Tr}^i \right\|_p^p, \quad (1)$$

where $\boldsymbol{\Theta}$ is the parameters of CNN; $N$ the number of training samples; $f(\cdot)$, a mapping function (network output); $p$, a p-norm (SRCNN [4] uses $p$=2); $D^T$, an up-sampling operator; $\boldsymbol{Y}_{Tr}^i$, an $i$-th training LR image (e.g., T2WI), and $\boldsymbol{X}_{Tr}^i$, the corresponding $i$-th training HR image (using external HR images). The guided SR includes the corresponding $i$-th training HR guidance image $\boldsymbol{X}_{TrG}^i$ (e.g., T1WI) to the network as follows:

$$L_{sg}(\boldsymbol{\Theta}) = \frac{1}{N}\sum_{i=1}^{N}\left\| f\left(D^T \boldsymbol{Y}_{Tr}^i, \boldsymbol{X}_{TrG}^i; \boldsymbol{\Theta}\right) - \boldsymbol{X}_{Tr}^i \right\|_p^p, \quad (2)$$

On the other hand, unsupervised SR such as ZSSR [8] cannot use training HR images. Therefore, ZSSR uses test LR image$\boldsymbol{Y}_{Te}^i$ as the HR image and its simulated degradation as LR image $DB\boldsymbol{Y}_{Te}^i$ (using internal LR training image) as follows:

$$L_{zssr}(\boldsymbol{\Theta}) = \frac{1}{M}\sum_{i=1}^{M}\left\| f\left(D^T DB\boldsymbol{Y}_{Te}^i; \boldsymbol{\Theta}\right) - \boldsymbol{Y}_{Te}^i \right\|_p^p, \quad (3)$$

where $B$ is a blurring operator; $D$, a down-sampling operator; and $M$, the number of training samples (data augmentation samples using test LR image). In the proposed method, we used the external LR images $\boldsymbol{Y}_{Tr}^i$ as HR images and its simulated degradation images as LR images $DB\boldsymbol{Y}_{Tr}^i$ (using external LR training images) with training degraded guidance images $DB\boldsymbol{X}_{TrG}^i$ as follows:

$$L(\boldsymbol{\Theta}) = \frac{1}{N}\sum_{i=1}^{N}\left\| f\left(D^T DB\boldsymbol{Y}_{Tr}^i, DB\boldsymbol{X}_{TrG}^i; \boldsymbol{\Theta}\right) - \boldsymbol{Y}_{Tr}^i \right\|_p^p, \quad (4)$$

In the test phase, we use the trained network for estimating HR image $\widehat{\boldsymbol{X}}_{Te}$ with HR guidance image $\boldsymbol{X}_{TeG}$ as follows:

$$\widehat{\boldsymbol{X}}_{Te} = f\left(D^T \boldsymbol{Y}_{Te}, \boldsymbol{X}_{TeG}; \boldsymbol{\Theta}\right) \quad (5)$$

### D. Cascade SR strategy and post-processing

In particular, during the unsupervised approaches (internal learning or external learning), when the magnification factor is larger, the simulated degradation LR images substantially lose their image features, and the ambiguity increases, thereby causing difficulties in estimation. Therefore, to obtain better results, we gradually increased the resolution with $r$ stage intermediate magnification scale $\sqrt[r]{s}$ for a magnification factor $s$ (e.g., when the magnification scale $s$ = 2 and the number of stage $r$ = 3, the enlarged size from input size ($H \times W$) are stage 1: $\sqrt[3]{2}H \times \sqrt[3]{2}W$, stage 2: $\sqrt[3]{2^2}H \times \sqrt[3]{2^2}W$ and stage 3: $2H \times 2W$). In this experiment, we set $r$ to 3. We trained three models with different scales. Firstly, we train the model using HR images that are $H \times W$ in size and LR images that are down-sampled in $1/\sqrt[3]{2}H \times 1/\sqrt[3]{2}W$ size. After the first training the learned model generates image with $\sqrt[3]{2}H \times \sqrt[3]{2}W$ dimensions. The generated $\sqrt[3]{2}H \times \sqrt[3]{2}W$ sized images are used as the HR image in the following phase, and its down-sampled $H \times W$ sized images are used as the LR image to train the second model. Now the model generates image with $\sqrt[3]{2^2}H \times \sqrt[3]{2^2}W$ dimensions. The third model is trained using $\sqrt[3]{2^2}H \times \sqrt[3]{2^2}W$ sized HR images and $\sqrt[3]{2}H \times \sqrt[3]{2}W$ sized down-sample LR images. During inference we generate HR images using these three models sequentially. This idea also used in other self-similarity-based SR methods [8]-[10] for natural images to improve the image quality. It also helps to prepare the training dataset by setting a small intermediate magnification factor. Although the methods retrain the network using intermediate estimated SR images, the quality of the final SR image is not guaranteed due to possible accumulated errors in the intermediate estimated images. On the other hand, our proposed method can estimate the SR image more precisely using the HR guidance image of each intermediate step. Errors can be mitigated by adopting guided SR for model learning.

In post-processing, we have used the iterative back projection (IBP) method [10] [27] to maintain the SR consistency by minimizing the errors between the estimated SR image projected onto the LR image space and input LR image. IBP is calculated by using the following equation:

$$\mathbf{X}_{t+1} = \mathbf{X}_t + B^{\mathrm{T}} D^T (\mathbf{Y} - DB\mathbf{X}_t), \quad (6)$$

where $\mathbf{X}_t$ is the estimated HR image after the $t$-th iteration; $\mathbf{Y}$, the observed LR image; $D$, the down-sampling operator; $D^{\mathrm{T}}$, the up-sampling operator. Further, $B$ and $B^{\mathrm{T}}$ are the blurring operators.

## III. Experiments

### A. Data preparation and training parameters

We have used NAMIC brain multimodality dataset [28], which includes 20 cases of publicly available brain MRI. The dataset has been acquired using a 3T GE at Brigham and Women's Hospital in Boston, MA, USA. The voxel dimensions of T1WI (TR = 7.4 ms, TE = 3 ms) and T2WI (TR = 2500 ms, TE = 80 ms) are 1 mm × 1 mm × 1 mm. The matrix size of both images is 256 × 256 × 176. The dataset is already registered. In this experiment, we used axial T1WI as the HR guidance image and axial T2WI as the LR image. To evaluate the SR performance, we generate LR T2WI, which is down-sampled with scaling factors 2 and 4 in the axial plane after Gaussian filtering with sigma set to $s/2\sqrt{2\ln 2}$ (the full width at half maximum equal to selected slice width [16, 29]). The Gaussian filter is widely used to generate the LR image as a blur operator [16, 36-38]. Although we can apply the SR method for volume with slice-by-slice processing, self-similarity-based SR methods such as ZSSR require large computational time for training and test. Thus, the center slices of each case are evaluated quantitatively with the peak signal-to-noise ratio (PSNR) and structural similarity (SSIM) [30]. The PSNR is defined as follows:

$$PSNR = 20\log_{10}\left(\frac{MAX}{\sqrt{MSE}}\right), \quad (7)$$

$$MSE = \frac{1}{HW}\sum_{i=0}^{H-1}\sum_{j=0}^{W-1}\left(\mathbf{X}(i,j) - \widehat{\mathbf{X}}(i,j)\right)^2, \quad (8)$$

where $MSE$ is the mean square error, $H$ and $W$ are the height and width of the image, $\mathbf{X}$ and $\widehat{\mathbf{X}}$ are the ground truth and estimated image, respectively, and $MAX$ is the maximum possible value of $\mathbf{X}$.

$\mathrm{SSIM_I}(x,y)$ for pixel ($x$, $y$) is calculated based on its

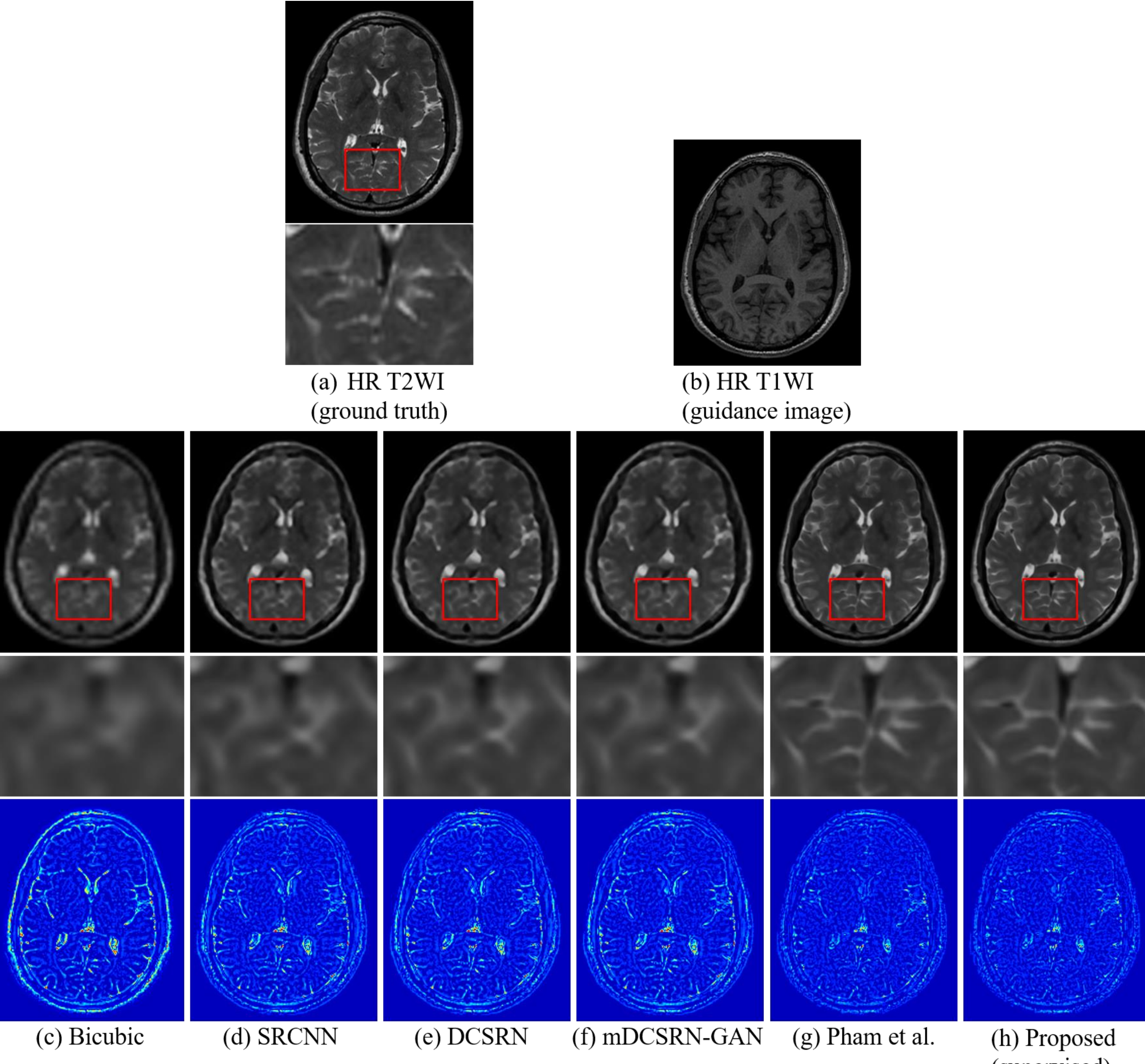

Fig. 3 Visualization of experimental results (top), the enlarged image within the red rectangle (middle), and the difference between target HR T2WI (ground truth) and each result (bottom) with a magnification factor of 4 in supervised approach. (a) Target HR T2WI (ground truth), (b) HR TIWI (guidance image), (c) Bicubic interpolation, (d) SRCNN [4], (e) 2D DCSRN [14], (f) 2D mDCSRN – GAN (b4u4) [15], (g) 2D Pham *et al.* (10L-ReCNN) [16], (h) Proposed (supervised + guide).

surrounding local patch, which is defined as follows:

$$SSIM_l(x,y) = \frac{(2\mu_X(x,y)\mu_{\hat{X}}(x,y)+c_1)(2\sigma_{X\hat{X}}(x,y)+c_2)}{(\mu_X(x,y)^2+\mu_{\hat{X}}(x,y)^2+c_1)(\sigma_X(x,y)^2+\sigma_{\hat{X}}(x,y)^2+c_2)}, \quad (9)$$

where $\mu_X$ and $\sigma_X$ are the Gaussian weighted mean and standard deviation of the local patch in $\mathbf{X}$, $\mu_{\hat{X}}$ and $\sigma_{\hat{X}}$ are the Gaussian weighted mean and standard deviations of the local patch in $\hat{\boldsymbol{X}}$, respectively; $\sigma_{X\hat{X}}$ is the Gaussian weighted covariance of the two local patches, $c_1 = (0.01 \cdot L)^2$, $c_2=(0.03 \cdot L)^2$, and $L$ is the dynamic range of the image.

The SSIM of the image is the average of $SSIM_l$, which is expressed as follows:

$$SSIM = \frac{1}{HW}\sum_{i=0}^{H-1}\sum_{j=0}^{W-1} SSIM_l(i,j) \quad (10)$$

We implemented all our models in TensorFlow [31] on a workstation with NVIDIA GeForce GTX 1080 Ti GPU (11GB memory). We used $L_1$ loss ($p = 1$) with an Adam optimizer [32]. The initial learning rate was set to $10^{-3}$, and the rate was divided by 10 if the loss did not continuously reduce over 10 steps. We stopped the training when the learning rate became less than $10^{-6}$. We have used image patches for training the networks. Image patches with size $64 \times 64$ are used in supervised methods and size $48 \times 48$ are used in unsupervised methods. A batch size of 32 is used for training. Other 3D models [14]-[16] were reimplemented by converting three-dimensional (3D) convolutional layers into two-dimensional (2D) models (i.e., from $3 \times 3 \times 3$ to $3 \times 3$) in our experiments. For data augmentation, we generated images by flipping horizontally, rotating 0, 90, 180, and 270 degrees in supervised methods. For in ZSSR and the proposed unsupervised methods (Model 1, 2, and Proposed methods), we rotated images at 15- and 30-degree intervals, applied horizontally flipping, and rescaling to reduce the resolution from scaling factors 0.5–1.0 in intervals of 0.1 and 0.2. We employed an IBP method which is adopted only for one iteration in all SR methods. The calculation time of the proposed unsupervised SR method was about 68 ms and the time of the IBP method (adopted for one iteration) was only about 1 ms per image. We used four-fold cross-validation to evaluate the proposed methods using external learning. The number of original training samples is 2640 (176 images $\times$ 15 subjects) and the number of test samples is 5 (1 image × 5 subjects) in each fold. We validated 20 test images. We used the cascade SR strategy by setting $r$ to 3 for both scaling factors 2× and 4× in the unsupervised SR (Table II), and applied a Gaussian filter as an intermediate blurring operator. Its standard deviation was set to $\sqrt[r]{s}/2\sqrt{2\lambda \ln 2}$, where $\lambda$ was set to 2. Following the [16], we did not perform any bias correction or image denoising preprocessing on the NAMIC dataset.

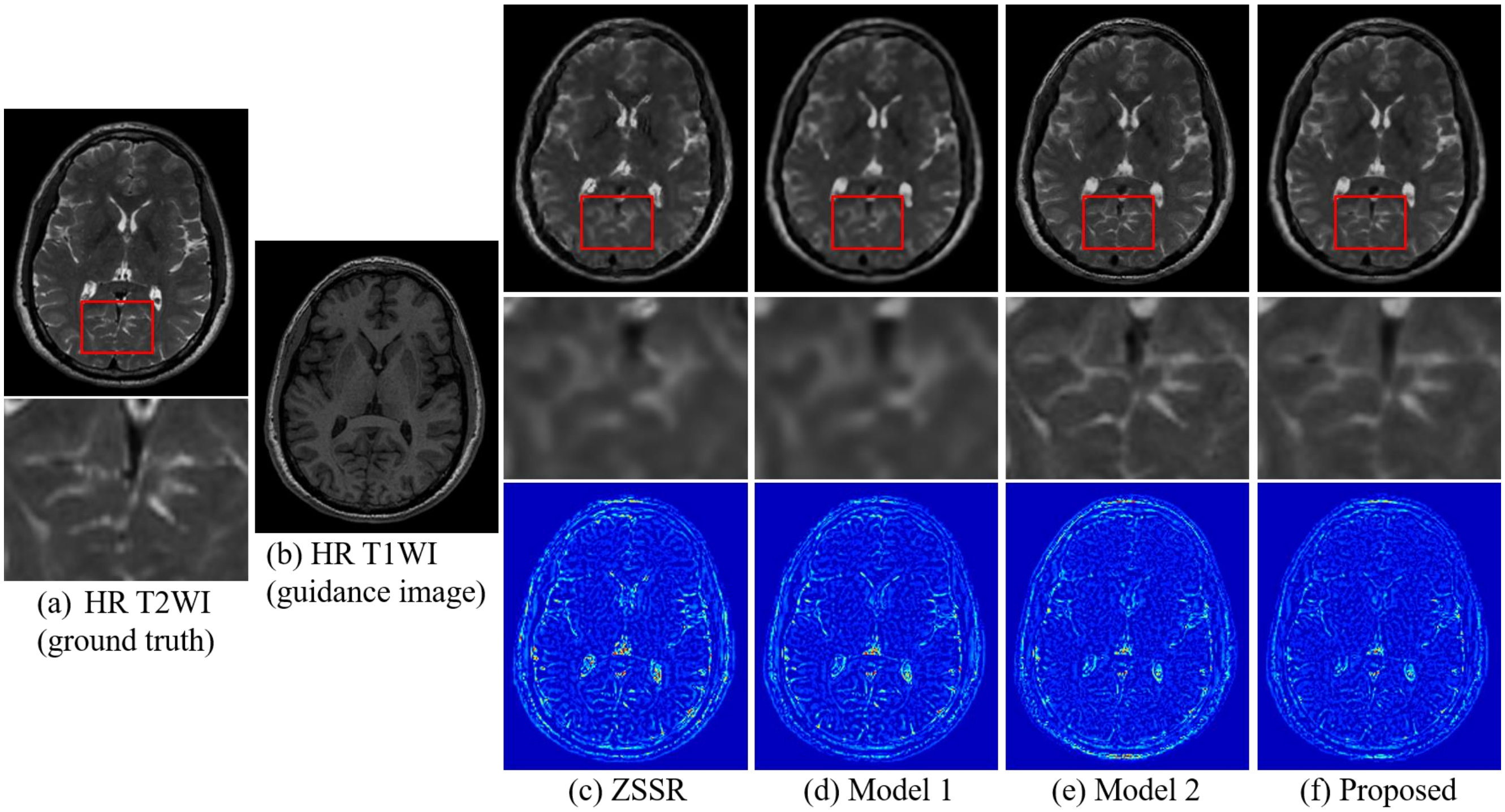

Fig. 4 Visualization of experimental results (up), the enlarged image within the red rectangle (middle), and the difference between target HR T2WI (ground truth) and each result (down) with a magnification factor of 4 in unsupervised approach. (a) Target HR T2WI (ground truth), (b) HR TIWI (guidance image), (c) ZSSR [8], (d) Model 2 (unsupervised external-learning), (e) Model 3 (unsupervised internal-learning + guide), (f) Proposed (unsupervised external-learning + guide).

TABLE I
Quantitative evaluation with average PSNR and SSIM [30] values of 20 test images in the supervised learning using external HR training images

| Method | Guide | 2x | | 4x | |
|---|---|---|---|---|---|
| | | PSNR | SSIM | PSNR | SSIM |
| Bicubic interpolation | | 34.55 | 0.961 | 29.62 | 0.872 |
| SRCNN [4] | | 38.15 | 0.981 | 31.58 | 0.918 |
| 2D DCSRN [14] | | 38.66 | 0.984 | 31.73 | 0.922 |
| 2D mDCSRN – GAN (b4u4) [15] | | 38.55 | 0.983 | 31.69 | 0.921 |
| 2D Pham *et al.* (10L-ReCNN) [16] | ✓ | 40.93 | 0.989 | 35.31 | 0.964 |
| Proposed (supervised) | ✓ | **41.48** | **0.990** | **36.04** | **0.969** |

TABLE II
Quantitative evaluation with average PSNR and SSIM [30] values of 20 test images in the unsupervised learning using internal LR training images or external LR training images

| Method | Internal LR | External LR | Guide | 2x | | 4x | |
|---|---|---|---|---|---|---|---|
| | | | | PSNR | SSIM | PSNR | SSIM |
| ZSSR [8] | ✓ | | | 37.97 | 0.982 | 30.71 | 0.908 |
| Model 1 | | ✓ | | 38.94 | 0.984 | 32.10 | 0.928 |
| Model 2 | ✓ | | ✓ | 39.13 | 0.985 | 32.60 | 0.943 |
| Proposed (unsupervised) | | ✓ | ✓ | **39.90** | **0.987** | **34.39** | **0.957** |

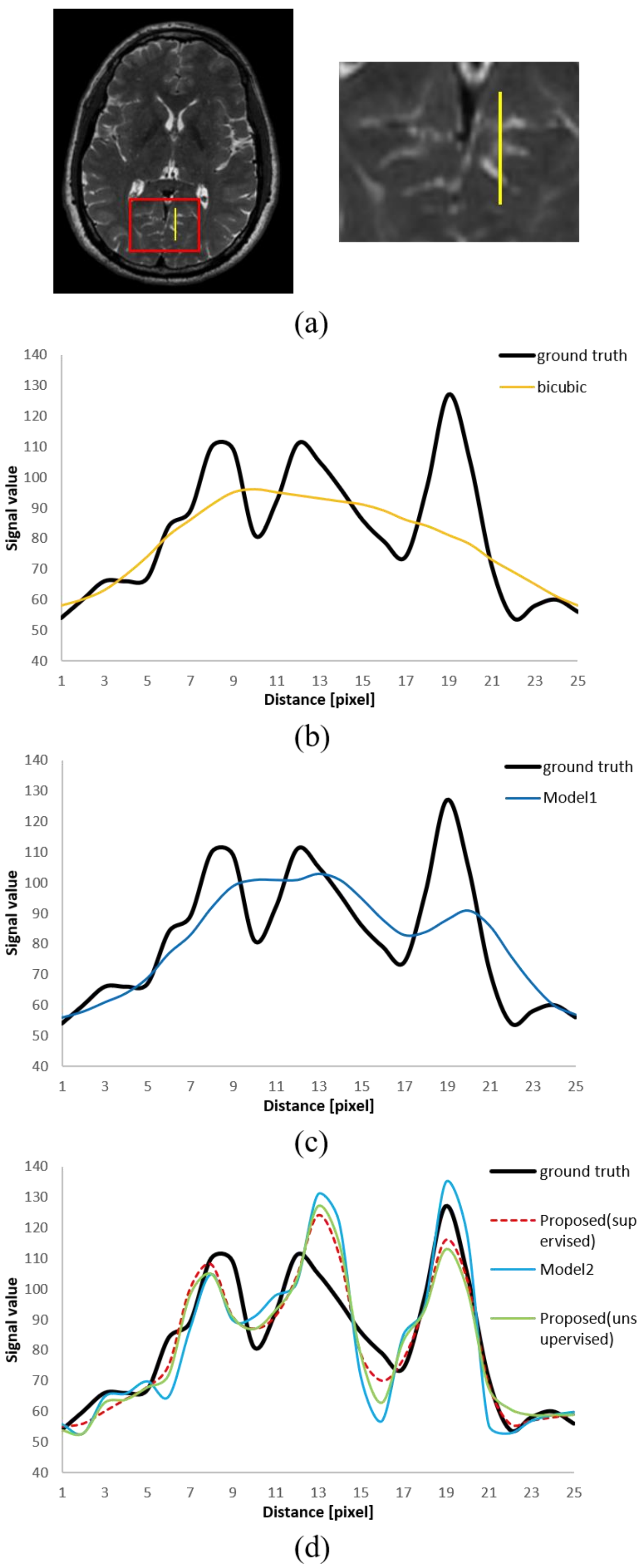

Fig. 5 Line profiles of a small portion marked by yellow line in (a).

### *B. Results and discussion*

First, we conducted comparative experiments using external HR and LR image pairs for training as the ideal case (i.e., general supervised learning). We compared the proposed network of Fig. 2 to state-of-the-art methods, including three methods without the guided HR image [4] [14] [15] and one method with the guided image [16]. The results are summarized in Table I. The values of PSNR and SSIM are an average of 20 test images. As shown in the table, proposed (supervised) method and [16] (supervised) that use the guided HR image outperform that do not use the guided HR image. Compared to [16], the proposed method (Fig.2) achieved better results in both 2 × and 4× cases. We found that all SR methods outperformed the conventional bicubic interpolation method. Thereafter, we conducted unsupervised learning experiments. The results are summarized in Table II. We compared the proposed method (represented as “Proposed (unsupervised)”) with the state-of-the-art unsupervised method ZSSR [8], which uses internal LR training images. In order to validate the effect of key components (external learning and guided image) of the proposed method, we also performed ablation study, which are represented as “Model 1” and “Model 2”. The model 1 represents the proposed network using external LR training images. Both ZSSR and model 1 do not use the guided HR image. Model 2 and proposed (unsupervised) method employ the proposed network of Fig. 2 directed with the guided HR image using the internal and external LR images, respectively. As shown in Table II, performance is significantly improved using external LR training images and further improvement is achieved using guided HR image. The proposed method achieved the best results in both 2× and 4× cases in unsupervised approach.

The supervised SR results from different methods for one subject with a magnification factor of 4 are shown in Figs. 3(c) to 3(h). The unsupervised SR results are shown in Figs. 4(c) to 4(f). The first and second rows are the reconstructed images and their enlarged small portions (marked by the red rectangle in the first row), respectively. The images in the third row are difference between the target HR T2WI (ground truth) and each result. Figures 3(a), 3(b), 4(a) and 4(b) show the HR T2WI image (ground truth) and its corresponding HR T1WI (guided image), respectively. Figure 3(c) shows the result obtained by conventional bicubic interpolation, and Figs. 3(d) to 3(h) show the supervised SR results obtained by the SRCNN [4], 2D DCSRN [14], 2D mDCSRN – GAN (b4u4) [15], 2D Pham *et al.* (10L-ReCNN) [16], and proposed (supervised + guide) methods, respectively. Figures 4(c) to 4(f) show the unsupervised results obtained by ZSSR [8], model 1 (external LR), model 2 (internal LR+ guide), and proposed method (external LR + guide), respectively. As shown in the enlarged images, the reconstructed image is not very clear even with the external HR training images (supervised learning) if we do not use the guided HR T1WI image. The quality of the reconstructed images can be improved significantly using the guided T1WI image even when only using the LR training images (i.e., unsupervised learning).

Table III Quantitative evaluation with average PSNR and SSIM values of 20 test images by the different iteration numbers in the cascade strategy.

| Cascade [times] | PSNR/SSIM |
|---|---|
| 1 | 32.11 / 0.936 |
| 2 | 34.14 / 0.955 |
| 3 | 34.39 / 0.957 |

Table IV Comparison of different loss with average PSNR and SSIM values of 20 test images.

| Loss | PSNR/SSIM |
|---|---|
| L1 | 34.39 / 0.957 |
| L2 | 34.29 / 0.956 |

Table V Quantitative evaluation with average PSNR and SSIM values of 20 test images by the different registration errors.

| Registration error | PSNR/SSIM |
|---|---|
| 0 mm | 34.39 / 0.957 |
| 1 mm | 32.96 / 0.941 |
| 2 mm | 32.42 / 0.933 |
| 3 mm | 32.29 / 0.931 |
| ZSSR [8] | 30.71 / 0.908 |

To facilitate a more detailed comparison, we also show the line profiles of the small portion. A yellow line is marked in the Fig. 5(a). As shown in Fig. 5(b), the line profile of the ground truth (HR image) includes three peaks with a distance of four pixels between the first and second peaks, and a distance of seven pixels between the second and third peaks. In contrast, the line profile of bicubic interpolation (LR image) has only one peak due to the low-resolution (the resolution of the LR image is 1/4 of the HR image). However, the spatial resolution of the LR image can be improved by SR techniques, as shown in Figs. 5(c) and 5(d). Figure 5(c) shows the line profile of SR without the guided image (Model 1). As shown, SR without guided image can reconstruct two peaks (the second and third peaks) with greater distance (seven pixels), while the first peak cannot be separated from the second peak because it has a smaller distance (four pixels) to the second peak. The SR results (Model 2, and the proposed methods (supervised and unsupervised)) with the guided image are shown in Fig. 5(d). Now, all three methods reconstruct three peaks well, that demonstrates the effectiveness of the guided image. Our proposed (supervised) method achieves the best results among other three methods. Images of model 2 (unsupervised internal LR + HR guided) appears to be slightly over-reconstructed. As shown in Fig. 5(d), the proposed supervised and unsupervised methods both provide SR images of comparable quality. From the above results and analysis, we consider that SR with the guided image can improve the spatial resolution significantly. In addition, unsupervised SR with the external LR is more effective than the existing "zero-shot" SR [8], which only uses the internal LR.

We investigated the effect of cascade strategy with a magnification factor of 4. The results are summarized in Table III. As shown in Table III, the cascade model with three iterations obtained better PSNR and SSIM values than the model with fewer iterations. Therefore, the cascade strategy is effective for improving the image quality.

In recent year, L1 loss is widely used for SR reconstruction instead of L2 loss because of its higher performance [7]. We compared the effectiveness of L1 and L2 loss with a magnification factor of 4. The results are summarized in Table IV. According to the results, L1 loss obtained slightly higher PSNR and SSIM values than L2 loss.

The accuracy of registration may have an impact on the proposed approach which uses the guidance image. Therefore, we investigated the influence of registration error. To perform the experiments, we constructed three types of HR guidance datasets by slightly shifting the x, y and z direction by 1 mm, 2 mm and 3mm, respectively. The volumes are generated by randomly shifting the x, y and z axes in the following direction: -1 or 1mm, -2 or 2 mm, and -3 or 3 mm respectively. We trained the model using the dataset with registration error and evaluated the test images. The results are summarized in Table V. The accuracy decreases proportionately to the registration error. However, compared to the existing unsupervised method (i.e., ZSSR [8]), the proposed approach obtained higher PSNR and SSIM values even with a registration error of 3 mm.

## IV. Conclusion

We introduced a novel external-learning-based unsupervised multimodal priors guided SR for the resolution enhancement of MRI images (T2WI). Our proposed method can directly utilize accumulated enormous medical image resources at the daily clinical site. Further, as the proposed method does not require HR training images for SR, it can be used with a variety of medical imaging modalities for which it is challenging to get HR images for training. We also demonstrate the effectiveness of guided SR directed by multimodal priors (HR T1WI) for constructing the detailed anatomical structure and its advantages over the state-of-the-art method. In future work, we aim to tackle the challenge of building an accurate and efficient unsupervised 3D model. In addition, instead of using the iterative back projection method for each test image, we will use deep internal learning after the external learning to refine the SR results.

## References

[1] N. Gordillo, E. Montseny, and P. Sobrevilla, "State of the art survey on MRI brain tumor segmentation," *Magn. Reson. Imaging*, vol. 31, no. 8, pp. 1426-1438, Oct. 2013.

[2] D. García-Lorenzo, S. Francis, S. Narayanan, D. L. Arnold, and D. L. Collins, "Review of automatic segmentation methods of multiple sclerosis white matter lesions on conventional magnetic resonance imaging," *Med. Image Anal.*, vol. 17, no.1, pp. 1-18, Jan. 2013.

[3] I. Despotović, B. Goossens, and W. Philips, "MRI Segmentation of the Human Brain: Challenges, Methods, and

Applications," *Comput. Math. Method Med.*, vol. 2015, no. 450341, pp. 1-23, 2015.

[4] C. Dong, C. C. Loy, K. He, and X. Tang, "Image Super-Resolution Using Deep Convolutional Networks," *IEEE Trans. on Pattern Anal. Mach. Intell.*, vol. 38, no. 2, pp. 295-307, Feb. 2015.

[5] J. Kim, J. K. Lee, and K. M. Lee, "Accurate Image Super-Resolution Using Very Deep Convolutional Networks," in *Proc. IEEE Conf. on Comput. Vis. Pattern Recognit. (CVPR)*, Las Vegas, NV, USA, Dec. 2016, pp. 1646-1654.

[6] C. Ledig, L. Theis, F. Huszár, J. Caballero, A. Cunningham, A. Acosta, A. Aitken, A. Tejani, J. Totz, Z. Wang, and W. Shi, "Photo-Realistic Single Image Super-Resolution Using a Generative Adversarial Network," in *Proc. IEEE Conf. on Comput. Vis. Pattern Recognit. (CVPR)*, Honolulu, HI, USA, Nov. 2017, pp. 4681-4690.

[7] Y. Zhang, Y. Tian, Y. Kong, B. Zhong, and Y. Fu, "Residual Dense Network for Image Super-Resolution," in *Proc. IEEE Conf. on Comput. Vis. Pattern Recognit. (CVPR),* Salt Lake City, UT, USA, Jun. 2018, pp. 2472-2481.

[8] A. Shocher, N. Cohen, and M. Irani, ""Zero-Shot" Super-Resolution Using Deep Internal Learning," in *Proc. IEEE Conf. on Comput. Vis. Pattern Recognit. (CVPR),* Salt Lake City, UT, USA, Dec. 2018, pp. 3118-3126.

[9] D. Glasner, S. Bagon, and M. Irani, "Super-Resolution Form a Single Image" in *Proc. IEEE 12th Int. Conf. on Comput. Vis.*, Kyoto, Japan, Sep./Oct. 2009, pp. 349-356.

[10] R. Timofte, R. Rothe, and L. V. Gool, "Seven ways to improve example-based single image super resolution" in *Proc. IEEE Conf. on Comput. Vis. Pattern Recognit. (CVPR)*, Las Vegas, NV, USA, Jun. 2016, pp. 1865-1873.

[11] Y. Iwamoto, X. H. Han, S. Sasatani, K. Taniguchi,W. Xiong, and Y. W. Chen, "Super-Resolution of MR Volumetric Images Using Sparse Representation and Self-Similarity" in *Proc. 21st Int. Conf. on Pattern Recognit. (ICPR)*, Tsukuba, Japan, Nov. 2012, pp. 3758-3761.

[12] A. Jog, A. Carass, and J. L. Prince, "Self Super-Resolution for Magnetic Resonance Images" in *Proc. Int. Conf. Med. Image Comput. Comput. Assit. Intervention (MICCAI)*, Athens, Greece, Oct. 2016, pp. 553-560.

[13] C. Zhao, A. Carass, B. E. Dewey, and J. L. Prince, "Self Super-Resolution for Magnetic Resonance Images Using Deep Networks" in *Proc. IEEE 15th Int. Symp. on Biomed. Imaging*, Washington, DC, USA, Apr. 2018, pp. 365-368.

[14] Y. Chen, Y. Xie, Z. Zhou, F. Shi, A. G. Christodoulou, and D. Li, "Brain MRI Super Resolution Using 3D Deep Densely Connected Neural Networks" In *IEEE 15th Int. Symp. on Biomed. Imaging*, Washington, DC, USA, Apr. 2018, pp. 739-742.

[15] Y. Chen, F. Shi, A. G. Christodoulou, Y. Xie, Z. Zhou, and D. Li, "Efficient and Accurate MRI Super-Resolution Using a Generative Adversarial Network and 3D Multi-Level Densely Connected Network," In *Proc. Int. Conf. on Med. Image Comput. Comput. Assist. Intervention (MICCAI)*, Granada, Spain, Sep. 2018, pp. 91-99.

[16] C. H. Pham, C. Tor-Díez, H. Meunier, N. Bednarek, R. Fablet, N. Passat, and F. Rousseau, "Multiscale brain MRI super-resolution using deep 3D convolutional networks," *Comput. Med. Imaging and Grap.*, vol. 77, Oct. 2019.

[17] F. Rousseau, "A non-local approach for image super-resolution using intermodality priors," *Med. Image Anal.*, vol. 14, no. 4, pp. 594-605, Aug. 2010.

[18] J. V. Manjón, P. Coupé, A. Buades, D. L. Collins, and M. Robles, "MRI Superresolution Using Self-Similarity and Image Priors," *Int. J. Biomed. Imaging*, vol. 2010, Dec. 2010.

[19] Y. Iwamoto, X. H. Han, A. Shiino, and Y. W. Chen, "Fast super-resolution with iterative-guided back projection for 3D MR images," in *Proc. SPIE 10574, Med. Imaging 2018*, Houston, Texas, USA, Mar. 2018.

[20] Y. Wei, Q. Yuan, H. Shen, and L. Zhang, "Boosting the Accuracy of Multispectral Image Pansharpening by Learning a Deep Residual Network," *IEEE Geosci. Remote. Sens. Lett.*, vol. 14, no. 10, pp. 1795-1799, Aug. 2017.

[21] C. Guo, C. Li, J. Guo, R. Cong, H. Fu, and P. Han, "Hierarchical Features Driven Residual Learning for Depth Map Super-Resolution" *IEEE Trans. on Image Process.*, vol. 28, no. 5, pp. 2545-2557, Dec. 2019.

[22] K. Takeda, Y. Iwamoto, and Y. W. Chen, "Color Guided Depth Map Super-Resolution based on a Deep Self-Learning Approach," in *Proc. IEEE Conf. on Consum. Electron. (ICCE)*, Las Vegas, NV, USA, Jan. 2020.

[23] K. He,X. Zhang, S. Ren, and J. Sun, "Deep Residual Learning for Image Recognition," in *Proc. IEEE Conf. on Comput. Vis. Pattern Recognit. (CVPR)*, Las Vegas, NV, USA, Dec. 2016, pp. 770-778.

[24] G. Huang, Z. Liu, L. V. D. Maaten, and K. Q. Weinberger, "Densely Connected Convolutional Networks," in *Proc. IEEE Conf. on Comput. Vis. Pattern Recognit. (CVPR)*, Honolulu, HI, USA, Nov. 2017, pp. 4700-4708.

[25] S. Ioffe, and C. Szegedy, "Batch Normalization: Accelerating Deep Network Training by Reducing Internal Covariate Shift," in *Proc. 32nd Int. Conf. on Mach. Learn. (ICML)*, Lille, France, Jul. 2015, pp. 448-456.

[26] X. Glorot, A. Bordes, and Y. Bengio, "Deep Sparse Rectifier Neural Networks," in *Proc. 14th Int. Conf. on Artif. Intell. Stat. (AISTATS)*, Fort Lauderdale, FL, USA, Apr. 2011, pp. 315-323.

[27] M. Irani, and S. Peleg, "Improving Resolution by Image Registration," *CVGIP: Graph. Models Image Process.*, vol. 53, no. 3, pp. 231-239, May 1991.

[28] NAMIC: Brain Multimodality dataset, http://hdl.handle.net/1926/1687, last accessed 2020/3/4.

[29] H. Greenspan, G. Oz, N. Kiryati, and S. Peled, "MRI inter-slice reconstruction using super-resolution," *Magn. Reson. imaging*, vol. 20, no. 5, pp. 437-446, Jun. 2002.

[30] Z. Wang, A. C. Bovik, H. R. Sheikh, and E.P. Simoncelli, "Image Quality Assessment: From Error Visibility to Structural Similarity," *IEEE Trans. on Image Process.*, vol. 13, no.4, pp. 600-612, Apr. 2004.

[31] M. Abadi, A. Agarwal, P. Barham, E. Brevdo, Z. Chen, C. Citro, G. S. Corrado, A. Davis, J. Dean, M. Devin, S. Ghemawat, I. Goodfellow, A. Harp, G. Irving, M. Isard, R. Jozefowicz, Y. Jia, L. Kaiser, M. Kudlur, J. Levenberg, D. Mané, M. Schuster, R. Monga, S. Moore, D. Murray, C. Olah, J. Shlens, B. Steiner, I. Sutskever, K. Talwar, P. Tucker, V. Vanhoucke, V. Vasudevan, F. Viégas, O. Vinyals, P. Warden, M. Wattenberg, M. Wicke, Y. Yu, and X. Zheng, "TensorFlow: Large-Scale Machine Learning on Heterogeneous Distributed Systems," 2015. Software available from tensorflow.org.

[32] D. P. Kingma, and J. Ba, "Adam: A Method for Stochastic Optimization," in *Proc. Int. Conf. on Learn. Represent. (ICLR)*, San Diego, CA, USA, May 2015.

[33] Jiqing Huang, Lihui Wang, Jin Qin, Yi Chen, Xinyu Cheng, and Yuemin Zhu, "Super-Resolution of Intravoxel Incoherent

Motion Imaging Based on Multisimilarity," *IEEE Sens. J.,* Vol. 20, No. 18, pp. 10963-10973, 2020.
[34] Yongbing Zhang, Siyuan Liu, Chao Dong, Xinfeng Zhang, and Yuan Yuan, "Multiple Cycle-in-Cycle Generative Adversarial Networks for Unsupervised Image Super-Resolution," *IEEE Trans. on Image Process.*, Vol. 29, pp. 1101-1112, 2020.
[35] Vineeta Das, Samarendra Dandapat, and Prabin Kumar Bora, "Unsupervised Super-Resolution of OCT Images Using Generative Adversarial Network for Improved Age-Related Macular Degeneration Diagnosis," *IEEE Sens. J.*, Vol. 20, No. 15, pp. 8746-8756, 2020.
[36] Andrea Rueda, Norberto Malpica, and Eduardo Romero, "Single-image super-resolution of brain MR images using overcomplete dictionaries," *Med. Image Anal.* , 17, pp. 113-132, 2013.
[37] Feng Shi, Jian Cheng, Li Wang, Pew-Thian Yap, and Dinggang Shen, "LRTV: MR Image Super-Resolution With Low-Rank and Total Variation Regularizations," *IEEE Trans. on Med. Imaging*, Vol. 34, No. 12, 2015.
[38] Chi-Hieu Pham, "Deep Learning for medical image super resolution and segmentation," *Image Processing [eess.IV]. Ecole nationale supérieure Mines-Télécom Atlantique*, 2018.
[39] Yulun Zhang, Kai Li, Kunpeng Li, and Yun Fu, "MR Image Super-Resolution With Squeeze and Excitation Reasoning Attention Network," *in Proc. IEEE Conf. on Comput. Vis. Pattern Recognit. (CVPR)*, pp. 13425-13434, 2021.
[40] Shaocong Mo, Ming Cai, Lanfen Lin, Ruofeng Tong, Qingqing Chen, Fang Wang, Hongjie Hu, Yutaro Iwamoto, Xian-Hua Han, and Yen-Wei Chen, "Mutual Information-Based Graph Co-Attention Networks for Multimodal Prior-Guided Magnetic Resonance Imaging Segmentation," *IEEE Trans. on Circuits Syst. Video Technol.,* Vol. 32, No. 5, pp. 2512-2526, 2022.

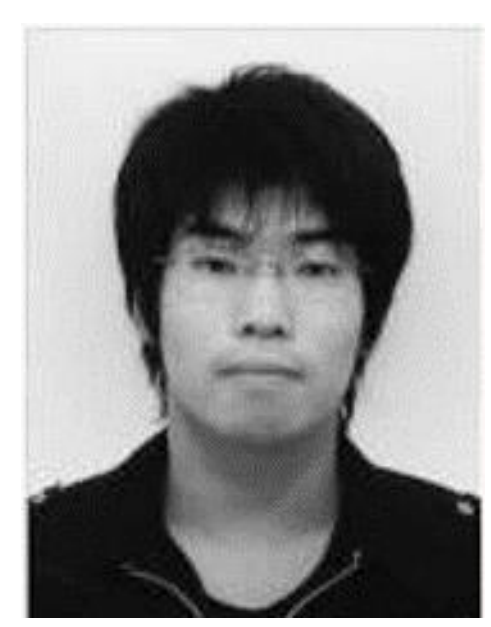

**Yutaro Iwamoto** received the B.E. and M.E., and D.E. degree from Ritsumeikan University, Shiga, Japan in 2011 and 2013, and 2017, respectively.

He is currently an Assistant Professor at Osaka Electro-Communication University, Osaka, Japan. His current research interests include medical image processing and computer vision, and deep learning.

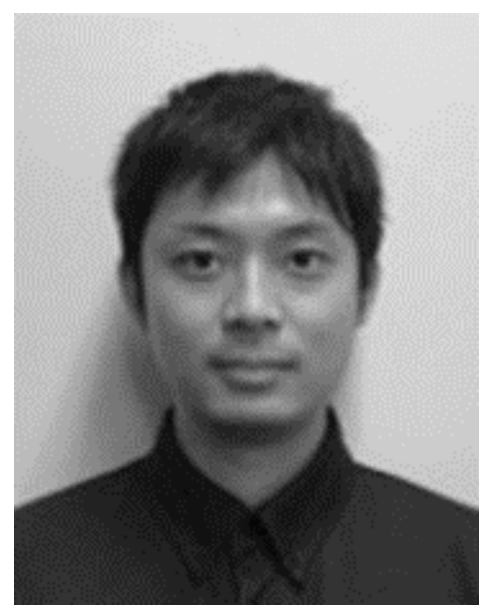

**Kyohei Takeda** received the B.E. and M.E. degree from Ritsumeikan University, Shiga, Japan in 2017 and 2020, respectively.

His current research interests include medical image processing and deep learning.

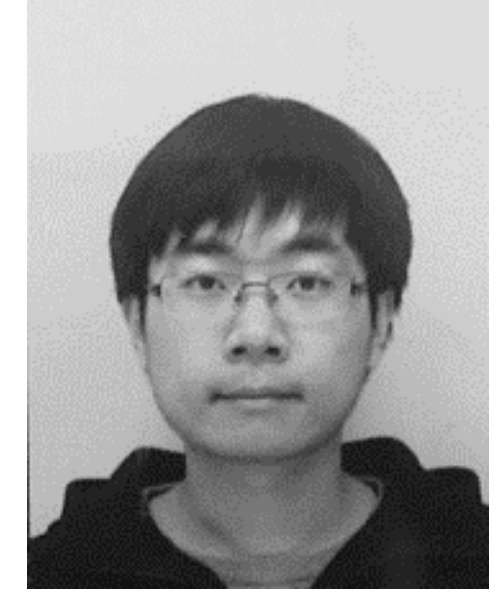

**Yinhao Li** received the B.E. degree from Southeast University Chengxian College, Nanjing, China, in 2013, and the M.E. and D.E. degrees from Ritsumeikan University, Kusatsu, Japan, in 2018 and 2021, respectively. He was a Postdoctoral Fellow with the Zhejiang Lab, China, from 2021 to 2022. He is currently an Assistant Professor with the College of Information Science and Engineering, Ritsumeikan University, Japan. His research interests include computer vision, medical image analysis, and deep learning. He was a recipient of the IEEE CE East Japan Chapter ICCE Young Scientist Paper Award in 2018, and Japan Society for the Promotion of Science (JSPS) Research Fellowship for Young Scientists in 2019.

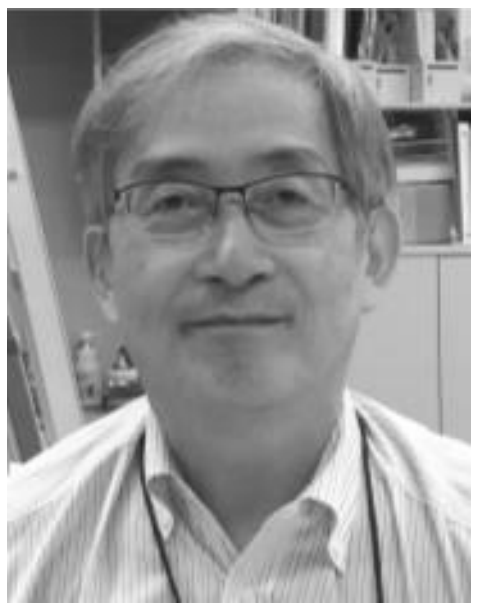

**Akihiko Shiino** received the M.D. from Shiga University of Medical Science, Shiga, Japan in 1983, and Ph.D. in 1988. From 1992-1993, he was a research fellow at University of Pennsylvania (Department of Physical Chemistry, and Radiologic Physics). From 1983-1999, he was a faculty at Shiga University of Medical Science. He is currently an associate professor at the Department of surgery, Shiga University of Medical Science, more than 30 years experience in clinical practice as a board certified neurosurgeon and neuroscientist.

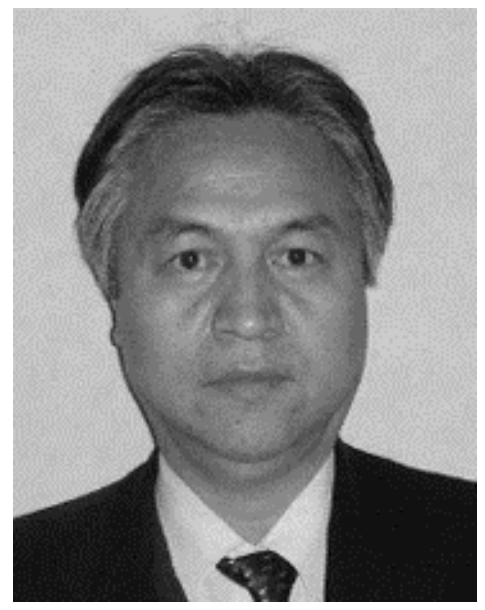

**Yen-Wei Chen** received a B.E. degree in 1985 from Kobe Univ., Kobe, Japan, a M.E. degree in 1987, and a D.E. degree in 1990, both from Osaka University, Osaka, Japan. From 1991 to 1994, he was a research fellow with the Institute for Laser Technology, Osaka. From October 1994 to March 2004, he was an associate Professor and a professor with the Department of Electrical and Electronic Engineering, University of the Ryukyus, Okinawa, Japan. He is currently a professor with the college of Information Science and Engineering, Ritsumeikan University, Kyoto, Japan. He is also a visiting professor with the Zhejiang Lab, China and the College of Computer Science and Technology, Zhejiang University, China. He is an associate Editor of International Journal of Image and Graphics (IJIG) and an associate Editor of the International Journal of Knowledge based and Intelligent Engineering Systems. His research interests include pattern recognition, image processing and machine learning. He has published more than 200 research papers in these fields.